\documentclass[preprint,aps,floatfix,showpacs,preprintnumbers,
]{revtex4}
\usepackage{amssymb}
\usepackage{epsfig}
\usepackage{graphicx}
\usepackage{dcolumn}
\usepackage{bm}

\begin{document}

\title{Complete spin extraction from semiconductors near
ferromagnet-semiconductor interfaces}
\author{V.~V.~Osipov$^{1,2}$, V.~N.~Smelyanskiy$^{2}$, and A.~G.~Petukhov$%
^{3}$}
\date{\today }
\affiliation{$^{1}$New Physics Devices, LLC, 2041 Rosecrans Avenue, 
El Segundo, CA  90245
\\
$^{2}$NASA Ames Research Center, Mail Stop 269-3, Moffett Field, CA 94035 \\
$^{3}$Physics Department, South Dakota School of Mines and Technology, Rapid
City, SD 57701 }

\begin{abstract}
We show that spin polarization of electrons in nonmagnetic semiconductors
near specially tailored ferromagnet-semiconductor junctions can achieve
100\%. This effect is realized even at moderate spin injection coefficients
of the contact when these coefficients only weakly depend  on the current.
The effect of complete spin extraction occurs at relatively strong electric
fields and arises from a reduction of spin penetration length due to the
drift of electrons from a semiconductor towards the spin-selective tunnel
junction.
\end{abstract}

\maketitle

Combining carrier spin as a new degree of freedom with the established
bandgap engineering of modern devices offers exciting opportunities for new
functionality and performance. This new field of semiconductor physics is
referred to as semiconductor spintronics \cite{Zut,Aw}. The injection of
spin-polarized electrons into nonmagnetic semiconductors (NS) is of
particular interest because of the relatively large spin-coherence lifetime, 
$\tau _{s}$, and the promise for applications in both ultrafast low-power
electronic devices \cite{Zut,Aw,Datta,Hot,OBO} and in quantum information
processing \cite{Aw,QIP,QC}. The main challenge is to achieve a high spin
polarization, $P_{n}$, of electrons in NS. The characteristics of the
spintronic devices dramatically improve when $P_{n}\rightarrow 100\%$.

It has been concluded in all previous theoretical works on spin injection 
\cite{Aron,Mark,Rash,Flat,Alb,BO,OB} that $P_{n}$ cannot exceed 
either the spin polarization of the carriers in the spin source or
the spin injection coefficient, $\gamma $, of the ferromagnet-semiconductor
junction \cite{Remk}. This conclusion does not contradict existing
experiments in which different magnetic materials such as magnetic
semiconductors, half-metallic ferromagnets, and ferromagnetic metals (FM)
have been used as spin sources \cite{Zut,Aw}. FM are widely used in
semiconductor technology. The Curie tempeartures of these materials are
usually much higher than the room temperature. The greatest value of $%
P_{n}\simeq $ 32\%, was achieved for Fe-based junctions \cite{Jonk,Ohno}
with approximately the same polarization of the source.

One of the obstacles for the spin injection from FM into NS is a high and
wide Schottky barrier that usually forms at the metal-semiconductor
interfaces \cite{sze}. The spin injection corresponds to a reverse current
of the Schottky FM-S junction. This current is usually extremely small \cite%
{sze}. Therefore, a thin heavily doped $n^{+}$-S layer between FM and S must
be used to increase the current \cite{sze,Jonk,Alb,OB,BO}. This layer
greatly reduces the thickness of the barrier and increases its tunneling
transparency. The greatest values of $P_{n}$ were found in such FM- $n^{+}$-$%
n$-S structures \cite{Jonk}.

Thus, the spin injection is the tunneling of spin polarized electrons from
FM into NS in reverse-biased FM-S structures. Since the tunneling is a
symmetric process the spin selective transport must also occur in the
forward-biased junctions when electrons are emitted from NS into FM \cite{BO}%
. In these junctions the electrons with a certain spin projection can be
efficiently extracted from NS while the opposite spin electrons will
accumulate in NS near FM-S interface \cite{BO}. Spin extraction from NS was
predicted by I. Zutic \textit{et al.} \cite{Zut1} for forward-biased p-n
junctions containing a magnetic semiconductor and was experimentally found
in forward-biased MnAs/GaAs Schottky junction \cite{Step}. However the
predicted and observed values of $P_{n}$ were rather small.

In this letter we demonstrate a possibility for achieving complete
spin polarization of electrons in NS near forward-biased FM-S junctions with
moderate spin injection coefficient, $\gamma $. The effect is based on spin
extraction and nonlinear dependence\ of the nonequilibrium spin density on
the electric field. We consider a FM-$n^{+}$-$n$-S structure containing a
heavily doped degenerate $n^{+}$-S layer, Fig.~1. We use a
standard assumption of spin injection\cite%
{Aron,Mark,Rash,Flat,Alb} that $\gamma$ of the FM-$n^{+}$-S contact only weakly
depends on the total current $J$ due to a high density of degenerate
electrons in the $n^{+}$-S layer (see below). In the forward-biased
structure unpolarized electrons drift from the bulk of NS to the contact.
Because of the spin selectivity of the contact the electrons with spin $%
\sigma =\uparrow $ (up-electrons) at $\gamma >0$ are extracted from NS, i.e. 
$\delta n_{\uparrow l}=(n_{\uparrow l}-n_{s}/2)<0$, and electrons with spin $%
\sigma =\downarrow $ (down-electrons) are accumulated, i.e. $\delta
n_{\downarrow }=(n_{\downarrow }-n_{s}/2)>0$, near the contact \cite{BO}.
Here $n_{s}$, $n_{\uparrow l}$ and $n_{\downarrow l}$ are the equilibrium
electron density in NS and densities of up-and down-electrons, respectively,
at the boundary between the $n^{+}$-S layer and high-resistant NS region ($%
x=l$ in Fig.~1(a)). The quantity $\left\vert \delta n_{\uparrow
l}\right\vert $ increases with the electric field, $E$ \cite{BO}.
In sufficiently strong fields, the drift efficiently compresses the spin
polarized electrons to the boundary. As a result \cite{Flat,BO}, the spin
penetration length $L$ decreases with the current $J$ [cf. white and red
curves in Fig.~1(a)]. Note, that due to $\delta n_{\downarrow }=-\delta
n_{\uparrow }$, the diffusion flow of up-electrons is directed along the
electron drift while the diffusion flow of down-electrons is in the opposite
direction, Fig.~1(a). The superlinear increase of the spin diffusion flows
with $J$ can be compensated only by an increase of the spin density $%
n_{\downarrow }$ up to $n_{s}$ and a decrease of $n_{\uparrow }$ down to
zero. In other words, spin polarization of the electrons in NS near FM- $%
n^{+}$-S contact $\left\vert P_{nl}\right\vert =\left\vert \delta
n_{\uparrow l}-\delta n_{\downarrow l}\right\vert /n_{s}=2\left\vert \delta
n_{\uparrow l}\right\vert /n_{s}$ can reach 100\% when the current is
sufficiently large.

\begin{figure}[htb]
\includegraphics[width=.77\linewidth,clip=true]{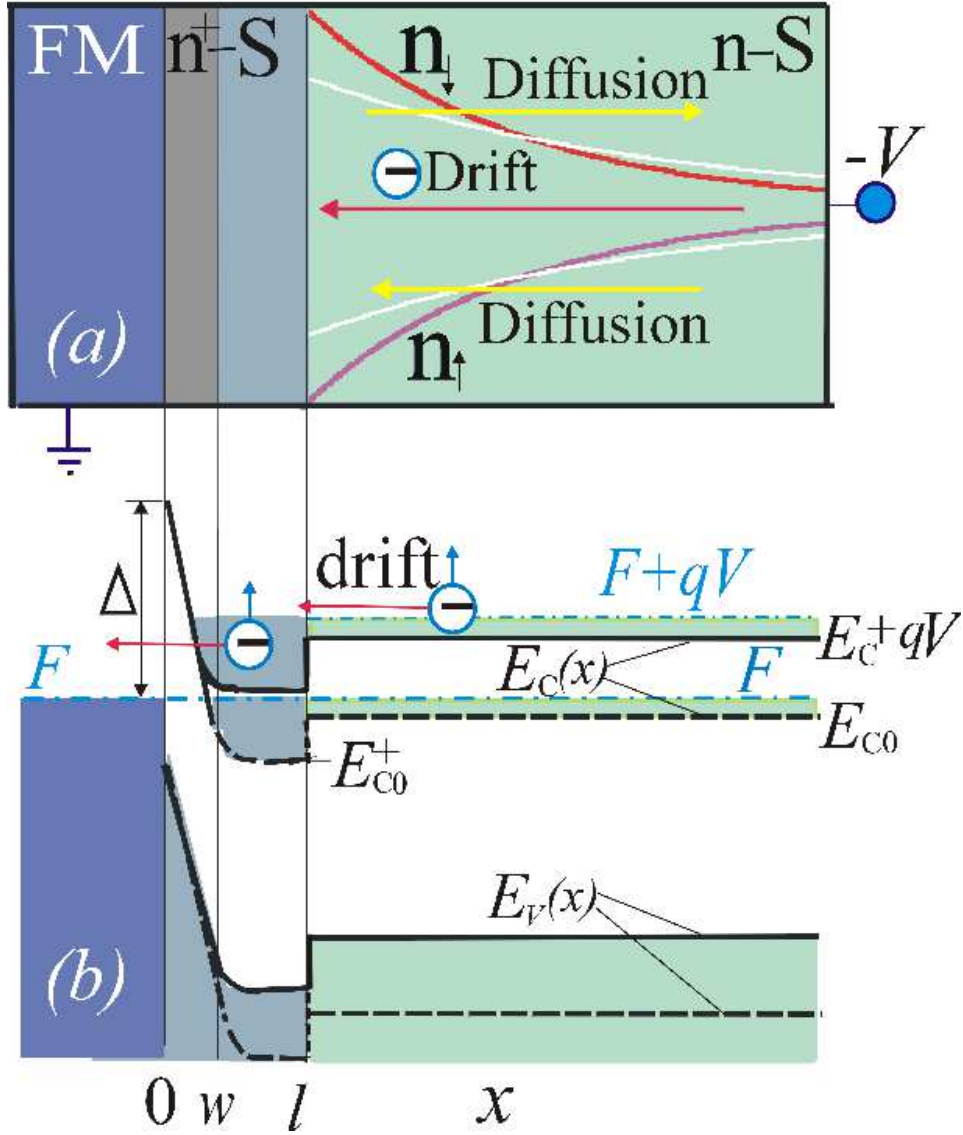}
\caption{{\bf(a)} 
FM-$n^{+}$-$n$-S heterostructures containing a thin heavily doped
degenerate semiconductor layer ($n^{+}$-S) sandwiched between the
ferromagnetic metal (FM) and donor doped degenerate nonmagnetic
semiconductor (NS) region ($n$-S). White and red curves display
spatial distributions of densities  $n_{\uparrow }$ and
$n_{\downarrow }$ at  small and large currents,
respectively; {\bf (b)} 
Energy diagrams of equilibrium (dashed
curves) and forward biased (solid curves)  FM-$n^{+}$-S junction 
for the case of a ``narrow''-bandgap $n^{+}$-S layer and a ``wide''-bandgap
$n$-S region. $F$ is the Fermi level; $w$ and $l$ are the
thicknesses of the Schottky barrier and  $%
n^{+}$-S layer, respectively; $E_{C}(x)$ and $E_{V}(x)$ are the bottom of the
conduction band and top of the valence band, respectively.}
\end{figure}

Let us consider for simplicity the case when the diffusion constant and
mobility of up- and down-electrons are the same constants: $D_{\uparrow
}=D_{\downarrow }=D$ and $\mu _{\uparrow }=$ $\mu _{\downarrow }=$ $\mu $.
This standard assumption \cite{Aron,Mark,Rash,Flat,Alb} is valid for
nondegenerate NS (the peculiarities of degenerate NS are
discussed below). In this case the currents of up- and down electrons with $%
\sigma =\uparrow ,\downarrow $ are given by the equations \cite%
{Aron,Flat,OB,BO} 
\begin{eqnarray}
J_{\sigma } &=&q\mu n_{\sigma }E+qD\frac{d\delta n_{\sigma }}{dx},
\label{eq} \\
dJ_{\uparrow }/dx &=&q(n_{\uparrow }-n_{\downarrow })/2\tau _{s}\text{ \ }
\label{Jx},
\end{eqnarray}%
where $q$ is the magnitude of the elementary charge.
It follows from conditions of the continuity of the total current and
electroneutrality that $J(x)=J_{\uparrow }+J_{\downarrow }=\mathrm{const,}$
and $n(x)=n_{\uparrow }+n_{\downarrow }=n_{s}=\mathrm{const}$. This means that $%
E(x)=J/q\mu n_{s}=\mathrm{const}$ and $\delta n_{\uparrow }(x)=-\delta
n_{\downarrow }(x)$. Then the solution of Eqs. (\ref{eq})-(\ref{Jx}) reads 
\cite{Aron,Flat,OB,BO} 
\begin{eqnarray}
\delta n_{\uparrow }(x) &=&P_{nl}\frac{n_{s}}{2}\exp [-(x-l)/L],\text{ }
\label{nx} \\
\text{where }L &=&(1/2)\left( \sqrt{4L_{s}^{2}+L_{E}^{2}}\pm L_{E}\right) 
\label{L}
\end{eqnarray}%
where $P_{nl}=P_{n}(l)=2\delta n_{\uparrow l}/n_{s}$ is the spin
polarization of the up-electrons at $x=l$ (Fig. 1), $L_{s}=\sqrt{D\tau _{s}}$
and $L_{E}=\mu \tau _{s}\left\vert E\right\vert =$ $L_{s}\left\vert
J\right\vert /J_{s}$ are the spin diffusion and drift lengths, respectively,
and $J_{s}=qn_{S}D/L_{s}$. The signs $\pm $ correspond to the reversed, $J<0$%
, and forward biases, $J>0$, respectively. From Eqs. (\ref{eq})-(\ref{nx})
we find that the currents at $x=l$ are 
\begin{equation}
J_{\uparrow l,\downarrow l}=\frac{J}{2}\pm J\frac{\delta n_{\uparrow l}}{%
n_{s}}\mp qD\frac{\delta n_{\uparrow l}}{L}=\frac{J}{2}\mp \frac{J_{s}}{2}%
\frac{L}{L_{s}}P_{nl}.  \label{Pn}
\end{equation}%
It follows from Eq.~(\ref{Pn}) that the electron spin polarizations, $%
P_{nl}=2\delta n_{\uparrow }/n_{s}$, and the spin injection coefficient, $%
\gamma _{l}=(J_{\uparrow l}-J_{\downarrow l})/J$, near the boundary are
related by the equation 
\begin{equation}
P_{nl}=-\gamma _{l}\frac{JL_{s}}{J_{s}L}=\frac{-2J\gamma _{l}}{\sqrt{%
(2J_{s})^{2}+J^{2}}\pm \left\vert J\right\vert }  \label{Pnl}
\end{equation}%
Thus, we see that for the case of the spin injection (reversed bias, sign +) 
$\left\vert P_{nl}\right\vert <\left\vert \gamma _{l}\right\vert $ in
accordance with previous works \cite{Aron,Flat,OB}. Another situation is
realized in \emph{the forward-biased FM-S junctions}, sign $-$ in Eq.~(\ref%
{Pnl}). Here the spin penetration depth $L$ (\ref{L}) decreases with the
current $J$ and according to (\ref{Pnl}) $\left\vert P_{nl}\right\vert $ 
\emph{approaches 1 (100\%) }when 
\begin{eqnarray}
J &\rightarrow &J_{t}\equiv J_{s}(\left\vert \gamma _{l}\right\vert +\gamma
_{l}^{2})^{-1/2}\text{ }  \label{thC} \\
\text{and }L &\rightarrow &L_{t}\equiv L_{s}\sqrt{\left\vert \gamma
_{l}\right\vert /(1+\left\vert \gamma _{l}\right\vert ).}  \label{thL}
\end{eqnarray}

In {\em degenerate} NS the diffusion constants depend on electron densities: $%
D_{\sigma }/\mu _{\sigma }=(D/\mu )(2n_{\sigma }/n_{s})^{2/3}$ at low
temperatures T$\ll \mu $. In this case we can find $E$ from Eqs. (\ref{eq})
and $J=J_{\uparrow }(x)+J_{\downarrow }(x)$. Then, substituting $E$ into Eq.
(\ref{eq}), we obtain $J_{\uparrow }$. Using this $J_{\uparrow }$ and 
Eq.~(\ref{Jx})
we find a diffusion-drift equation for $\delta n_{\downarrow }(x)$ with a
bi-spin diffusion constant, $D(P_{n})=(D/2)\left( 1-P_{n}^{2}\right) ^{2/3}%
\left[ (1+P_{n})^{1/3}+(1-P_{n})^{1/3}\right]$, which depends on $P_{n}=2\delta
n_{\downarrow }/n_{s}$. One can see that $D(P_{n})\rightarrow 0$ when $%
\left\vert P_{n}\right\vert \rightarrow 1$. It means that the effective spin
diffusion length $L_{s}(P_{n})=[D(P_{n})\tau _{s}]^{1/2}$ decreases with the
current because $\left\vert P_{nl}\right\vert \rightarrow 1$ near $x=l$.
Thus, an additional mechanism of a decrease of the spin penetration 
length $L$ with current $%
J$ occurs in a degenerate NS. As a result, the decay of $P_{n}(x)$ is sharper,
particularly near $x=l$, as shown in the inset in Fig. 2. Therefore, in
degenerate NS the condition of complete spin extraction, $\left\vert
P_{nl}\right\vert =1$, can be reached at lower threshold currents and
greater spin lengths as compared with those given by (\ref{thC}) and (\ref%
{thL}) for nondegenerate NS. For instance, numerical analysis shows that the
threshold values $J_{t}=1.3J_{s}$ and $L_{t}=0.37L_{s}$ at $\gamma _{l}=0.3$
for $D_{\sigma }/\mu _{\sigma }=(D/\mu )(2n_{\sigma }/n_{s})^{2/3}$while $%
J_{t}=1.6J_{s}$ and $L_{t}=0.48L_{s}$ for the case $D_{\sigma }/\mu _{\sigma
}=const$. The effect of complete spin extraction from a degenerate NS can be
illustrated based on spatial and current dependences of quasi-Fermi levels $%
F_{\uparrow }$ and $F_{\downarrow }$ for up- and down-electrons,
respectively (Fig.~2). Indeed, due to the spin extraction the difference
between $F_{\uparrow }$ and $F_{\downarrow }$ near the FM-$n^{+}$- S contact 
increases with the current. Therefore, the value $F_{\uparrow }$
can reach the bottom of the conduction  band  $E_{c}$ in NS at $x=l$ (Fig.2) at
the current $J=J_{t}$. This implies that $\Delta F_{\uparrow }=F_{\uparrow
}-F=-\mu _{s}$ at this point and $n_{\uparrow l}\propto (F-E_{c}+\Delta
F_{\downarrow })\longrightarrow 0$, $n_{\downarrow l}=(n_{s}-n_{\uparrow
l})\rightarrow n_{s}$, i.e. $\left\vert P_{nl}\right\vert \rightarrow 1$.
Here $\mu _{s}=F-E_{c}$ and $F$\ are the Fermi energy and the equilibrium 
Fermi level of
electrons in NS, respectively. 
\begin{figure}[tbh]
\includegraphics[width=.77\linewidth,clip=true]{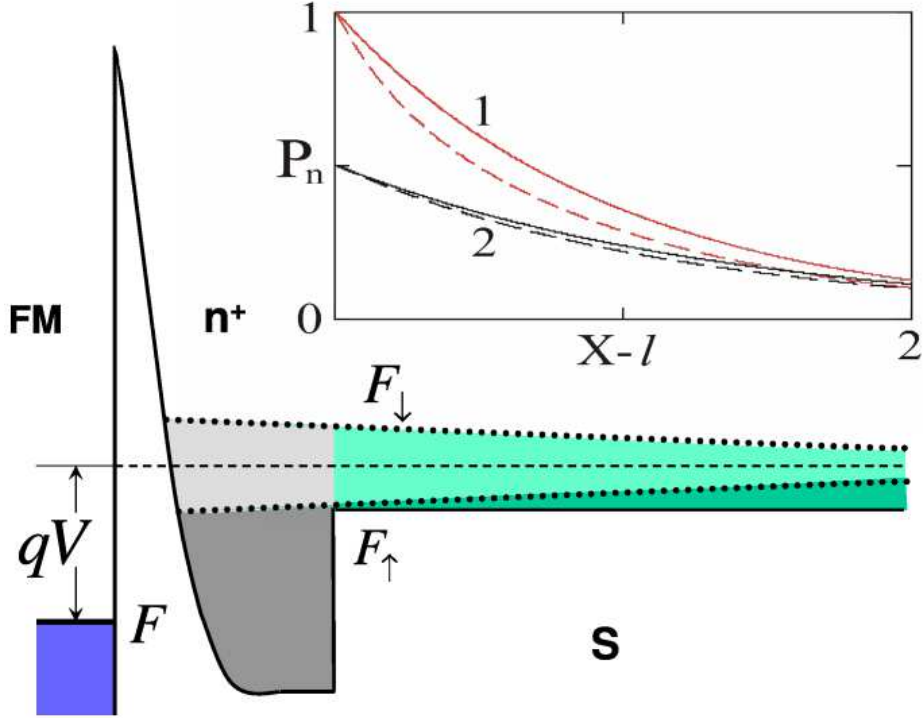}
\caption{Spatial dependences of the quasi-Fermi levels $F_{\uparrow }$ and $%
F_{\downarrow }$ for up- and down-electrons, respectively, 
at the threshold current $%
J=J_{t}$. The inset shows electron spin polarizations, $P_{n}(x)$ ($x$
is in units of $L_s$), for
cases $D_{\protect\sigma }/\protect\mu _{\protect\sigma }=const$ (solid
curves) and $D_{\protect\sigma }/\protect\mu _{\protect\sigma }=(D/\protect%
\mu )(2n_{\protect\sigma }/n_{s})^{2/3}$ (dashed curves) at $\protect\gamma %
=0.3$ and currents $J=1.6J_{s}$ (curves 1, solid line), $J=1.3J_{s}$ (curves 1,
dashed line),  and $J=0.8$ (curves 2). Dashed curves are the numerical solutions
of the diffusion-drift equation for $\protect\delta n_{\downarrow
}(x)$ with a bi-spin diffusion constant $D(P_{n})$ (see text).}
\end{figure}

In reality, however, our theory, which is based on the
consideration of two nonequilibrium ensembles  of the up- and down-electrons,
becomes invalid  when $n_{\uparrow l}\rightarrow 0$.
Our approach is justified only when the time of electron-electron
collisions within each of these systems is much less than $\tau _{s}$.
Moreover, at large currents $J>J_{t}$ the value of $\left\vert
P_{nl}\right\vert =2\left\vert \delta n_{\downarrow l}\right\vert
/n_{s}=\left\vert 2n_{\downarrow l}-n_{s}\right\vert /n_{s}$\ becomes
greater than 1 (see e.g. (\ref{Pnl})), i.e. spin density $n_{\downarrow l}$
at $x=l$ exceeds the equilibrium electron density, $n_{s}$. 
Therefore, the condition of local electroneutrality $n_{\uparrow l}+$ $%
n_{\downarrow l}=n_{s}$ is violated and a space charge arises near $x=l$ in
Fig.1. This charge will change the field $E(x)$ and the total electron
density in the vicinity of $x=l$. The complete set of equations consists 
of Eqs.~(\ref{eq})-(\ref%
{Jx}), $J=J_{\uparrow }(x)+J_{\downarrow }(x)=\mathrm{const}$, the and Poisson's
equation: $\varepsilon \varepsilon _{0}dE/dx=\rho$, where $\rho
=q(n_{s}-n_{\uparrow }+$ $n_{\downarrow })$ and $\varepsilon \varepsilon _{0}
$ is the permittivity of the NS. Our calculations for the case of $\gamma _{l}=const$
show that, as expected, the characteristic scale of the nonuniform-field
region is determined by a relatively short screening length and the value
of $\left\vert P_{nl}\right\vert $
in the degenerate NS\ is close to $1$ near $x=l$ at $J\simeq J_{t}$.

One can see from (\ref{thC}) and (\ref{thL}) that the spin injection
coefficient of FM-$n^{+}$-S contact, $\gamma _{l}$, determines the threshold
current, $J_{t}$, and spin penetration depth, $L_{t}$. However our main
finding that $\left\vert P_{nl}\right\vert \rightarrow 1$ at $J\rightarrow
J_{t}$ remains vlid at any reasonable value of $\gamma _{l}$. The only
required condition is a relatively weak dependence of $\gamma _{l}$ on $J$
(see \cite{Remark}). This  can be realized in a FM-$n^{+}$-S junction
containing a heavily doped $n^{+}$-S layer. The donor concentration, $%
N_{d}^{+}$, and thickness, $l$, of this layer must satisfy the following
conditions: $l\gtrsim 3w$ and $N_{d}^{+}w^{2}q^{2}\simeq 2\varepsilon
\varepsilon _{0}\Delta $, where $\Delta $ and $w$ are the height and width
of the depletion Schottky layer, Fig. 1. The electron gas has to be highly
degenerate in a certain part of the $n^{+}$-S layer contiguous $n$-S layer.
The transition between the $n^{+}$-S and $n$-S layers should have a
discontinuous jump $\Delta _{0}=(E_{c}-E_{c}^{+})$ shown in Fig.1(b). This
is realized when the $n^{+}$-S layer has \textbf{a} narrower energy bandgap
than that of the $n$-S region. A similar diagram can also be  realized when $%
n^{+}$-S and $n$-S regions are made of the same semiconductor, but an
additional, acceptor-doped, ultrathin  layer is formed between the $n^{+}$-S
and $n$-S regions. The acceptor concentration $N_{a}$ and thickness $l_{a}$
of this layer have to satisfy the conditions: $N_{a}l_{a}^{2}q^{2}\simeq
2\varepsilon \varepsilon _{0}\Delta _{0}$ and $l_{a}\ll l$.

To demonstrate the weak dependence of $\gamma _{l}$ upon  the current $J$ through 
FM-$n^{+}$-S junction we use the common assumption that the electron energy $E
$, spin $\sigma ,$ and the lateral component $\vec{k}_{\parallel }$ of the wave vector $%
\vec{k}$  are conserved during tunneling. Then the  current  density of electrons with spin $\sigma
=\uparrow ,\downarrow $ tunneling through the Schottky barrier, i.e. between 
the points $x=w$\ and $x=0$ in Fig. 1, can be
expressed as \cite{OB,BO}: 
\begin{equation}
J_{\sigma w}=\frac{q}{h}\int dE[f(E-F_{\sigma w}^{+})-f(E-F)]\int \frac{%
d^{2}k_{\parallel }}{(2\pi )^{2}}T_{k\sigma },  \label{GEJ}
\end{equation}%
where $f(E-F)$ is the Fermi function, $F$ the Fermi level in FM, $F_{\sigma
w}^{+}$ quasi-Fermi levels up- and down-electrons in $n^{+}$- S layer near
the FM-S interface ($x=w$ in Fig.1), and $T_{k\sigma }$ is the transmission
probability. We also assume that the temperature $T\ll \mu _{s}^{+}/k_B$, 
and $\mu _{s}^{+}=(F-E_{c0}^{+})$ is the Fermi energy of degenerate
equilibrium electrons of the $n^{+}$-S layer. In this case the
nonequilibrium density of the electrons with spin $\sigma $ at $x=w$ 
reads
\begin{equation}
n_{\sigma w}^{+}=\frac{n^{+}}{2(\mu _{s}^{+})^{3/2}}(F_{\sigma
w}^{+}-E^+_{c0}-qV)^{3/2}=\frac{n^{+}}{2}\left[ 1+\frac{\Delta F_{\sigma w}^{+}}{%
\mu _{s}^{+}}\right] ^{3/2},  \label{ncc}
\end{equation}%
where $n^{+}$ is the equilibrium electron density at $x=w$;\ $E_{c0}^{+}$
is the bottom of conduction band in the $n^{+}$-S
region in equilibrium,  $V$ is the  bias voltage, and $\Delta F_{\sigma
w}^{+}=(F_{\sigma w}^{+}-F-qV)$. Using the approximate expression for $%
T_{k\sigma }$ \cite{BO,OB}, and Eqs. (\ref{GEJ}) - (\ref{ncc}) at \ $T\ll
\mu _{s}/k_B$, $\left\vert qV\right\vert <\mu _{s}^{+}$ and $w\gtrsim 3l_{0}$ we
obtain 
\begin{equation}
J_{\sigma w}=j_{0}d_{\sigma }T_{0}(\mu _{s}^{+})^{-5/2}\left[ \left( \mu
_{s}^{+}+\Delta F_{\sigma w}^{+}\right) ^{5/2}-\left( \mu _{s}^{+}-qV\right)
^{5/2}\right]   \label{jw}
\end{equation}%
where $j_{0}=qn_{s}^{+}v_{F}\alpha _{0}$, $\alpha _{0}\simeq 0.96(\kappa
_{0}l)^{1/3}\simeq 1$ and $T_{0}=\exp \left[ -\eta w\frac{(\Delta -\mu
_{s}^{+}-qV)^{1/2}}{l_{0}\Delta ^{1/2}}\right] $ and $d_{\sigma
}=v_{F}v_{\sigma 0}/(v_{t0}^{2}+v_{\sigma 0}^{2})$ is the tunneling
transparency and the spin selection factor of  FM-$n^{+}$-S contact; $%
\eta \simeq 4/3$, $l_{0}=(\hbar ^{2}/2m_{\ast }\Delta )^{1/2}$ is a
tunneling length, $v_{t0}=\sqrt{2(\Delta -qV)/m_{\ast }}$, $v_{\sigma
0}=v_{\sigma }(F+qV)$ and $v_{F}=\sqrt{3\mu _{s}^{+}/m_{\ast }}$ are
velocities of electrons with spin\ $\sigma $ and the energies $F+qV$ and $%
\mu _{s}^{+}$ in  FM and  $n^{+}$-S regioons, respectively, and $m_{\ast }$
effective mass of electrons in $n^{+}$-S layer.

Let us consider the case when\ the thickness of the $n^{+}$-S layer $l\ll
L_{s}^{+}$, but $l\gtrsim 3w$. Here $L_{s}^{+}=\sqrt{D^{+}\tau _{s}^{+}}$ is
the spin diffusion length in the $n^{+}$-S layer. Due to the condition $l\ll
L_{s}^{+}$ the  quasi-Fermi levels, $F_{\uparrow }$ and $F_{\downarrow }$ and
the spin currents change very weakly in the $n^{+}$-S layer (Fig. 2).
Therefore we can put $J_{\sigma w}\simeq J_{\sigma l}$ and $\gamma
_{w}\simeq \gamma _{l}$.  We noticed above that in degenerate $n^{+}$-S $%
\left\vert \Delta F_{\sigma w}^{+}\right\vert \simeq \mu _{s}=(E_{c}-F)$ at $%
x=w$ when $J\rightarrow J_{t}$. Due to $n_{s}^{+}\gg n_{s}$ the value $\mu
_{s}^{+}\propto (n_{s}^{+})^{2/3}\gg \mu _{s}$, therefore we can neglect $%
\left\vert \Delta F_{\sigma w}^{+}\right\vert $ in Eq.(\ref{jw}) in
comparison with $\mu _{s}^{+}$ when $qV\simeq \mu _{s}^{+}$ at $J\simeq
J_{t} $. In this case we find that the spin injection coefficient, $\gamma
_{w}=(J_{\uparrow w}-J_{\downarrow w})/J,$ and the total current of the FM-S
junction are equal 
\begin{eqnarray}
\gamma _{0} &=&(d_{\uparrow }-d_{\downarrow })/(d_{\uparrow }+d_{\downarrow
})  \label{PF} \\
J &=&J_{0}T_{0}\left[ 1-\left( 1-qV/\mu _{s}^{+}\right) ^{5/2}\right] .
\label{CT}
\end{eqnarray}%
Here $J_{0}=(d_{\uparrow }+d_{\downarrow })j_{0}$ and $\gamma _{0}$ depend
weakly on $V$ and $J$ ($\gamma _{0}$ can increase with $V$ \cite{BO}).
We note that 
$J_{0}\propto n_{s}^{+}=N_{d}^{+}$ while $J_{t}\propto n_{s}=N_{d}$,
and, therefore $J_0\gg J_t$.
We see that $\gamma _{l}\simeq \gamma _{w}=\gamma _{0}$ in the
forward-biased FM-$n^{+}$-$n$-S structures when $L_{s}^{+}>l\gtrsim 3l_{D}$, 
$J_0\gg J_t$, and $%
J_{0}T_{0}\sim J_{t}$ at $qV\simeq \mu _{s}^{+}$. In other words we suppose
that Rashba's condition \cite{Rash} is valid for the FM-$n^{+}$-S junction
and therefore the spin injection coefficient $\gamma _{l}$ only weakly depends
on the current at $J\lesssim J_{t}$. In this case, as we have shown above, the
spin polarization of electrons in degenerate $n$-S region near the $n^{+}$-S
layer, $P_{nl}\rightarrow 100\%$ as $J\rightarrow J_{t}$.

In real ferromagnets the situation is much more complex. In FMs there are
spin-polarized heavy d-electrons and nonpolarized light s-electrons with
very involved energy spectrum. Nonetheless our conclusion about the weak
dependence of the spin injection coefficient \ $\gamma _{l}$ on the current
remains valid for any complex spectrum.\ This conclusion is based on the
fact that the perturbations of the quasi-Fermi levels in $n^{+}$-S layer are
small: $\Delta F_{\sigma 0}^{+}\ll qV\leq \mu _{s}^{+}$. The latter
inequality follows from a very large mismatch of the carrier concentrations
in the heavily doped $\ n^{+}$-S layer and  NS region with higher 
resistivity: $%
n_{s}/n_{s}^{+}=N_{d}/N_{d}^{+}\ll 1$.

In conclusion, we emphasize that we have demonstrated a possibility 
of achieving 100\% spin polarization in NS via electrical spin
extraction, using FM-S contacts with moderate spin injection coefficients
that weakly depend on the current. The highly spin-polarized electrons,
according to the results of Ref. \cite{Kaw}, can be efficiently utilized to
polarize nuclear spins in semiconductors. They can also be used to spin
polarize electrons on impurity centers or in quantum dots located near the
FM-S interface. These effects are important for spin-based quantum
information processing \cite{Aw,QIP,QC}. The considered FM-$n^{+}$-$n$-S
heterostructures and FM-$n^{+}$-S contacts can be used as very efficient
spin polarizers or spin filters in most of the spin devices proposed to date 
\cite{Zut,Aw,Datta,Hot,OBO}. In particular, such devices as spin-based
high-frequency spin-transistors, square law detectors, frequency
multipliers, magnetic sensors \cite{OBO},  spin-light
emitting diodes (spin-LEDs) \cite{Jonk,BORad}, and spin-resonant tunneling
diodes (spin-RTDs) \cite{Pet} can be modified to significantly enhance their
performance.




\begin{thebibliography}{99}
\bibitem{Remk} Note that $\gamma $ is denoted as $P_{J}$ in Refs. \cite%
{OBO,BO,OB} and is called the spin polarization of current in FM- S contacts.
\bibitem{Remark} This condition is not valid for tunneling FM-S junctions
containing a $\delta -$doped layer and nondegenerated semiconductor\ layer
near FM-S interface \cite{BO}.
\end{thebibliography}
\end{document}